\documentstyle{jfm}

\catcode`\œ=\active \gdefœ{\setbox0=\hbox{0}\hbox to\wd0{}}

\ifCUPmtlplainloaded \else
  \def\upi{\pi} 
\fi
\ifCUPmtlplainloaded

  \renewcommand{\simeq}{\approx}
\fi
\ifCUPmtlplainloaded
  \font\bit = mtmib10 at 10.5pt \skewchar\bit ='177  
\else
  \font\bit = cmmib10 \skewchar\bit ='177  
\fi
\ifCUPmtlplainloaded \else
  \font\tenbmi=cmmib10 at 10pt  \skewchar\tenbmi ='177
  \font\sevenbmi=cmmib10 at 7pt \skewchar\sevenbmi ='177
  \font\fivebmi=cmmib10 at 5pt  \skewchar\fivebmi ='177

  \newfam\bmifam
  \textfont\bmifam=\tenbmi
  \scriptfont\bmifam=\sevenbmi
  \scriptscriptfont\bmifam=\fivebmi
  
\fi
\newsavebox{\thalfbox}
\sbox{\thalfbox}{$\textstyle\frac{1}{2}$}

\newsavebox{\shalfbox}
\sbox{\shalfbox}{$\scriptstyle\frac{1}{2}$}

\newsavebox{\squartbox}
\sbox{\squartbox}{$\frac{1}{4}$} 

\newsavebox{\etbox}
\sbox{\etbox}{\boldmath$\eta$}

\newsavebox{\astrutbox}
\sbox{\astrutbox}{\rule[-5pt]{0pt}{20pt}}

\ifnfsstwo

\fi \ifnfssone
  \newmathalphabet{\mathit}
    \addtoversion{normal}{\mathit}{cmr}{m}{it}
    \addtoversion{bold}{\mathit}{cmr}{bx}{it}

\fi \ifoldfss

\fi

\mathchardef\varLambda="0103

\ifCUPmtlplainloaded
\else
\fi
\ifCUPmtlplainloaded
  \let\bcdot=\undefined
  \NewSymbolFont{bldsym}{mtbsy10}{'60}
  \NewMathSymbol{\bcdot}{2}{bldsym}{01}
\else
  \font\tenbms=cmbsy10          \skewchar\tenbms ='60
  \font\sevenbms=cmbsy10 at 7pt \skewchar\sevenbms ='60
  \font\fivebms=cmbsy10 at 5pt  \skewchar\fivebms ='60

  \newfam\bmsfam
  \textfont\bmsfam=\tenbms
  \scriptfont\bmsfam=\sevenbms
  \scriptscriptfont\bmsfam=\fivebms

  \edef\bsy{\hexnumber\bmsfam}
  \mathchardef\bnabla="0\bsy72
  \mathchardef\bcdotsymbol="0\bsy01
  \def\bcdot{\,\bcdotsymbol\,}
\fi
\title[Pattern formation in the Faraday instability]{Pattern
formation at the bi-critical point of the Faraday instability}

\author[C. Wagner, H.-W. M\"uller and K. Knorr]%
{C.\ns W\ls A\ls G\ls N\ls E\ls R$
^1$%
\thanks{Present address:  Laboratoire de Physique
Statistique, UMR CNRS 8550, Ecole Normale Sup\'{e}rieure, 24 rue
Lhomond, 75231 Paris Cedex 05, France.},\ns
H.\ls W.\ns M\ls \"U\ls L\ls L\ls E\ls R $^2$\break
\and K.\ns K\ls N\ls O\ls R\ls R$^1$}

\affiliation{$^1$Institut  f\"ur Technische Physik, Universit\"at
des Saarlandes Postfach 151150, D-66041 Saarbr\"ucken,
Germany\\[\affilskip] $^2$Max Planck Institut f\"ur
Polymerforschung, Ackermannweg 10, D-55128 Mainz, Germany}

\pubyear{2000}
 \volume{538}
 \pagerange{000--000}
\date{?? and in revised form ??}
\begin{document}

\maketitle

\begin{abstract}
We present measurements on parametrically driven surface waves
(Faraday waves) performed in the vicinity of a bi-critical point
in parameter space, where modes with harmonic and subharmonic
time dependence interact. The primary patterns are squares in the
subharmonic and hexagons in the harmonic regime. If the primary
instability is harmonic we observe a hysteretic
secondary transition from hexagons to squares without a
perceptible variation of the fundamental wavelength. The
transition is understood in terms of a set of coupled Landau
equations and related to other canonical examples of phase
transitions in nonlinear dissipative systems. Moreover, the
subharmonic-harmonic mode competition gives rise to a variety of
new superlattice states. These structures are interpreted as
mediator modes involved in the transition between patterns of
fourfold and sixfold rotational symmetry.

\end{abstract}

\section{Introduction}
The Faraday experiment has nowadays become a model system for
pattern formation  in hydrodynamic systems (\cite {Faraday31}).
For a review see \cite {Miles90} and \cite{Mueller98}. Standing
waves are generated on the liquid air interface in response to a
 time periodic gravity modulation.
 Under typical laboratory conditions and assuming
that the excitation acceleration is sinusoidal with $g(t)=g_0+a
\sin\Omega t$ these surface waves oscillate with twice the period
of the external drive  (\cite{Benjamin54}). This is a consequence
of the parametric drive mechanism and denoted here as the {\em
subharmonic} Faraday resonance. Surface waves synchronous
({\em harmonic}) with the drive can be generated, too. They have been
 observed first by adding a second frequency component to the
excitation signal (\cite {Edwards94}). Later on, following a
suggestion of  \cite {Kumar96}, harmonic  Faraday waves have also
been excited with the usual  single frequency drive
(\cite{Mueller97}). This, however, requires rather extreme
(parameter) conditions, namely thin fluid layers in combination
with  drive frequencies lower than some threshold $f_b$.
Increasing $f=\Omega/(2 \pi) $ beyond $f_b$  lets the Faraday
waves resonate with their usual subharmonic time dependence
(\cite{Wagner00}). For operating frequencies $f$ close to the
bi-critical value $f_b$ the harmonic and subharmonic modes
compete. Owing to the dispersion of surface waves different
frequencies imply different wavelengths. As a consequence
non-linear pattern formation is affected in a significant manner:
Subharmonic modes ($f<f_b$) form square patterns, harmonic modes
($f>f_b$) hexagons. The occurrence of a primary hexagonal surface
tiling of the subharmonic regime is generically due to a three
wave interaction. At elevated drive amplitude we observe a
transition towards a square pattern. This is similar to the
canonical hexagon-line transition in Rayleigh-B\'{e}nard convection,
which can be observed if "non-Boussinesq" effects become
significant (\cite {Walden81} and \cite {Ciliberto88}). A
transition from hexagons to squares has been found only recently
in the B\'{e}nard Marangoni instability (\cite {Nitschke95}, \cite
{Bestehorn96} and \cite {Eckert98}).

The measurements presented here give a comprehensive account of
our investigations on Faraday pattern selection in the vicinity
of the bi-critical point. Thereby the interaction between
harmonic and subharmonic modes of different wavelengths
gives rise to new resonant phenomena: superlattices with either
fourfold or sixfold rotational invariance. Though superlattices
are very common in solid state and surface physics, they have been
found on macroscopic scales only recently (\cite{Pampaloni97},
\cite {Kudrolli98}, \cite {Arbell98}, \cite{Wagner99} or
\cite{Wagner00}).

Within a cascade of secondary phase transitions
superlattices are found to mediate between the two incompatible
symmetry classes, of squares and hexagons.
For instance a primary subharmonic pattern with quadratic surface tiling
experiences a crossover to a hexagonal superlattice via two
quadratic superlattices (\cite{Wagner00}) with a prominent
displacive character in one or two lateral directions. After
passing a phase with a hexagonal superlattice the transition process
reaches a pure hexagonal pattern characterized by a single wavelength
and oscillating in synchronous response to the external drive.

For several of the observed transitions we are able to provide
explanations in terms of resonant amplitude equations for the
governing spatial modes. The structure of these equations is
simply based on symmetry and resonance arguments.
In spite of their simplicity these equations provide an
understanding of many remarkable features of the superlattices, in particular their displacive
character. This phenomenological approach
is certainly facilitated by the small number of experimental
control parameters. This is unlike earlier experiments
(\cite {Kudrolli98} and \cite {Arbell98} or \cite {Wagner99}),
which use a more complicated multiple frequency drive or
a viscoelastic fluid to drive the system into the
bi-critical situation. Thereby different kinds of superlattices
have been reported as well. But clearly, a larger number of
control parameters renders a theoretical understanding more
unwieldy and less intuitive. For the theoretical approach to superlattices
see e.g. \cite{Dionne97} and \cite{Silber99}.

\section{Experimental set-up}
\subsection{Vibration system and sample fluid}
\begin{figure}
  \caption{Sketch of the experimental set-up. See text for further
   explanation.}
  \label {setup}
\end{figure}

Figure~\ref{setup} shows a schematic diagram of the experimental
setup. Its heart is a large displacement shaker unit (V617 Gearing
\& Watson Electronics Ltd.) connected to a 4kW power amplifier.
The shaker supplies a maximum force of $4670 N$ and a peak-to-peak
elevation of $s_{max} = 54 mm$. Such a large displacement is
necessary to obtain a sufficient acceleration $a$ at lower
driving frequencies. The drive signal for the power amplifier is
synthesized by means of a digital-analog card installed in a
Pentium PC. The actual acceleration of the container is measured
with a piezo-electric device, the amplified signal of which is
routed to the PC for data acquisition.  Since the characteristics
of the shaker turned out to be rather non-linear at operation
frequencies below $f=\Omega/(2 \pi)<10$Hz a continuous control of
the excitation signal was necessary. To guarantee a sinusoidal
container acceleration  $a \sin{\Omega t}$ the recorded
accelerometer signal was decomposed into Fourier components. The
parasitic higher harmonics of $\Omega$ were eliminated by
admixing Fourier contributions with appropriate inverse phases to
the excitation signal. Their amplitudes were determined by a
proportional control loop. That way the power spectrum of the
accelerometer signal is made monochromatic with a purity of
$99\%$.

The cylindrical container for the sample liquid was machined out
of aluminum and was anodized black. To avoid pollution and temperature drifts within the
fluid, the container was sealed with a glass plate. The inner container
diameter was
$d = 290$mm, the depth $50$mm. Over a distance of  $12 mm$ from
the edges of the container the depth continuously increases from
zero to the bottom. This ''soft boundary condition'' with an
average angle of $30^\circ$ helps to minimize the generation of
parasitic meniscus waves. A meniscus under vertical vibration
always emits waves with the frequency $f$ of the external drive.
Since these waves have non vanishing amplitudes even at
subcritical drive amplitudes $a < a_c$ they blur the onset
detection. The beach like boundary
fullfill their purpose well, at least above 10Hz.

 The probe fluid was a  low viscosity  Silicon oil
(Dow Corning 200) with the manufacturer specifications of
kinematic viscosity $\nu = 5\times 10^{-6}m^2/s$, surface tension
$\sigma = 0.0194 N/m$ and density $\rho = 920 kg/m^3$ at our
working temperature $T = 25^\circ C$. A heating foil was mounted
on the outside of the container. By means of a temperature
controller the temperature measured by a PT-100 resistor (embedded
in the container body) was regulated by $\pm 0.1^\circ C$.

\subsection{Visualization technique}

To visualize the surface profile we used a  full frame CCD camera
(Hitachi KPF-1) situated above the fluid surface in the center of a ring consisting
of $120$ LED´s. The ring had a radius of $R=0.3m$ and
its distance from the fluid surface was $L=1.50m$. The camera was
synchronized to the excitation signal with
an exposure time of $1/256$ of the drive period.
It follows from geometrical optics that only surface elements with
a certain steepness reflect light into the camera.

For an evaluation of the spatial symmetry of the surface
deformation $\zeta(x,y)$ we relied on a Fourier technique. To
that end the recorded light intensity $I(x,y)$ of a video image
was convoluted with a Gaussian window function and processed by a
FFT algorithm. This yields the two dimensional spatial power
spectrum $P(\bf{k})$.  To determine the wavelength of the pattern
$P({\bf k})$ is azimuthally averaged by integrating over circles
with constant radius $|{\bf k}|=k$. The primary peak in the
resulting one-dimensional spectrum is fitted by a Gauss function
the center of which determines the fundamental wave number.
Clearly, the resolution of this procedure is limited by the number
of wavelengths in the container. This is especially the
case for subharmonic Faraday waves where the
uncertainty of  $\Delta k/k$ is about $10\%$.

Due to the nonlinear relationship between the surface elevation
$\zeta(x,y)$ and the recorded light intensity $I(x,y)$, the power
spectrum entails higher harmonics of the fundamental wave number,
even if the surface profile $\zeta(x,y)$ does not. Thus the
relation between $I(x,y)$ and $\eta(x,y)$ is generally too
complicated to allow a reconstruction of the surface profile. Nevertheless for simple
surface patterns (such as squares) we have solved this ''inverse
problem'' by the following method: Starting from an estimated
surface profile composed of a small number of spatial Fourier
modes, the light distribution of the expected video image  was
computed by means of a ray tracing algorithm. Then we adapted the
mode amplitudes and their relative phases such as to optimize the
agreement between the calculated and recorded video picture.

A reconstruction of the full time dependence of an oscillating
surface wave pattern was not possible with our equipment.
Nevertheless, the electronic shutter of the camera provides an
easy and very sensitive technique to discriminate subharmonic
frequency components in an otherwise harmonic time signal. This
is because a harmonic time dependence $\zeta^h(t)$ is invariant
under the symmetry operation $t\rightarrow
t+\frac{2\upi}{\Omega}$ implying a frequency spectrum of integer
multiples of $\Omega$ thus $\zeta^h(t)=\sum_n \zeta_n e^{i n
\Omega t}$. In contrast the subharmonic time signal transforms
after one drive period as $\zeta^s\rightarrow-\zeta^s$ enforcing a
Fourier representation in the form of $\zeta^s(t)=\sum_n \zeta_n
e^{i [(n+1/2) \Omega t]}$. Thus by triggering the camera shutter
with the drive frequency $\Omega$, video images with a harmonic
time dependence appear stationary, while those with subharmonic
frequency contributions flicker due to a slight optical asymmetry
between heaps and hollow of the deformed surface. Note however,
that this trigger technique does not allow to identify harmonic
frequency components in an otherwise subharmonic spectrum.

\section{The onset of the Faraday instability}
It is well known that the stability problem of a free liquid
surface under gravity modulation (Faraday instability) can be
approximately mapped to that of a parametrically driven pendulum
(\cite{Rayleigh83}, \cite{Benjamin54}). The primary
resonance of which occurs at twice the period of the drive ({\em
subharmonic response}). However, as first pointed out by
\cite{Kumar96} the Faraday instability may also appear in
synchronous resonance with the external drive, usually denoted as
the {\em harmonic} response. The conditions under which the
harmonic resonance preempts the subharmonic one have been worked
out in detail by \cite{Cerda97} and \cite {Mueller97} revealing
that low filling levels in combination with small drive
frequencies are necessary. In the present experiment we choose a
fill height of $h=0.7mm$, which is -- at the operation
frequencies of $6 < f < 8$Hz -- comparable to the viscous
penetration depth $\xi=\sqrt{2\nu/\Omega}\approx 0.5mm$. For the
fluid parameters at hand a linear stability analysis of the flat
surface state (according to the method of \cite{Kumar94}, which
assumes a laterally infinite system) reveals the location of the
bi-critical point at a drive frequency of $f_b=6.3$Hz.
Figure~\ref{instability} shows neutral stability diagrams (drive
amplitude $a$ vs. wave number $k$) for both situations $f<f_b$
and $f>f_b$.
\begin{figure}
   \vspace{1pc}
\caption{Neutral stability curves $a(k)$ computed for the
parameters of the sample fluid at a drive frequency a)$f=6.25$Hz
$<f_b$ and b)$f=7.25$Hz $>f_b$. In a) the primary resonance is
harmonic, in b) it is subharmonic. Regions where the flat surface
state is unstable are shaded. Horizontal lines denote the
thresholds for secondary and higher order transitions (for
details see text).}
  \label {instability}
\end{figure}
Experimentally the critical acceleration $a_c$ (absolute minimum
of the neutral stability diagram) has been determined by setting
up the system at  a constant frequency and ramping $a$
quasi-statically in steps of $0.2\%$ suspended by intervals of
$240$s. The onset amplitude $a_c$ was defined when the camera
detected the first light reflex (figure~\ref{onsetvalues}a). To
enhance the detection sensitivity the surface was illuminated by
a diffusive light source from the side rather than using the
dark-field technique described above. This is because the latter
method requires a minimum surface gradient of
$|\nabla\xi(x,y)|=\tan\alpha\approx0.1$ for the onset detection.
We estimate the accuracy of our threshold determination by
$0.5\%$.

Once a standing wave pattern had covered the whole surface
the fundamental frequency
of the surface oscillation was determined with the help of the
electronic shutter of the video camera. That way we located the
transition point at a
bi-critical frequency $f_b=6.5\pm 0.1$Hz. After these preliminary
measurements we switched back  to the dark-field illumination to
proceed with the spatial pattern analysis. The critical wave
numbers $k_c^h$ and respectively $k_c^s$ were determined by
Fourier transforming a surface image taken at a driving strength
of $\varepsilon=(a-a_c/a_c)  \approx3\%$ (figure~\ref{onsetvalues}b). The
operating prescription $k_c=k(\varepsilon\approx 3\%)$  for the
determination of the critical wave number is motivated by the
fact that we were unable to detect any change of the wave number $k$
by varying $\varepsilon$ (see also \cite {Wernet01}).
The experimental results for $a_c$ and $k_c$ as well the
bi-critical frequency $f_b$ are (found to be) in good agreement
with the theoretical predictions. For the critical acceleration
the discrepancy is less than $2\%$. Here the uncertainty is
mainly due to errors in the determination of the small fill
height $h$. For larger values of $h$ the agreement improves up to
$1\%$. For the critical wave number the discrepancy between
theory and experiment is better than $4\%$ in the harmonic case
but it increases up to  $10\%$ on the subharmonic regime. This is due
to the spatial resolution, which becomes worse at larger
wavelength.
\begin{figure}
   \vspace{1pc}
  \caption{ (a) Critical amplitude $a_c$ and (b) critical wave number
  $k_c$ for the onset of the Faraday instability drawn as a function
  of the drive frequency $f$.
  The bi-critical point $f=f_b$ is located where the harmonic ($f<f_b$) and
  subharmonic ($f>f_b$) thresholds intersect. Symbols mark experimental
  data points, lines the theoretical results for a laterally infinite
  fluid layer. Circles and dotted lines refer to the harmonic response,
  squares and solid lines to the subharmonic one.}
  \label {onsetvalues}
\end{figure}
Owing to the abrupt change of the response frequency at $f=f_b$
the wave number shows a discontinuous jump (see
figure~\ref{onsetvalues}b). The empiric ratio of the wave numbers
at $f=f_b$ is found to be
\begin{equation}
{\frac{k_c^h}{k_c^{s}}}|_{exp}=1.58 \pm 0.15,
\end{equation}
in agreement with the prediction of the linear stability theory
${\frac{k_c^h}{k_c^{s}}}|_{theo}= 1.59$.

\section{Overview over the phase diagram}
\begin{figure}
   \vspace{1pc}
\caption{Phase diagram of the observed patterns obtained by
quasi-statically ramping the driving force. The symbols mark the
observed transition points between the different patterns, the
lines are guides for the eyes. The spatial ordering of the
patterns is indicated by Roman numerals, the arabic letters s and
h denote the character of the time dependence being either purely
subharmonic or harmonic. With s+h we indicate patterns formed by
a interaction of subharmonic and harmonic modes. The thick line
separates harmonic from subharmonic and s+h regions. I: flat
surface; II: subharmonically oscillating squares (figure~
\ref{w2w2image} a,b); IIIa: $\sqrt 2 \times \sqrt 2$ superlattice
p2mg (figure~\ref{w2w2image}c,d); IIIb: $\sqrt 2 \times \sqrt 2$
superlattice c2mm (figure~\ref{w2w2image}e,f); within the
subregion above the dotted line the pattern is time
dependent and disordered (see figure~\ref{transient});
IV: $\sqrt 3 \times \sqrt 3$
superlattice, (only for $f<6.9$Hz stationary, figure~\ref{w3w3image});
V: harmonically oscillating hexagons (figure~\ref{hexagonstosquares}a);
VI: harmonically oscillating squares (figure~\ref{hexagonstosquares}d);
VII: $2 \times 2$ superlattice (figure~\ref{2x2image});
VIII: local instability and droplet ejection.}
\label {phasediagram}
\end{figure}
The phase diagram shown in figure~\ref{phasediagram} has been
obtained at various constant driving frequencies
$f=\Omega/(2\pi)$ while ramping the driving amplitude from
$\varepsilon=-1\%$ up to $25\%$ in steps of $0.2\%$. After each
increment the ramp was suspended for 240 seconds to give the
system time to relax. Then  a photo or - in the case of time
dependent patterns a video film - of the surface state was taken.
At the point where a  new spatial or temporal mode appeared or an
existing one died out, the actual acceleration was defined as the
transition boundary to a new "phase". At the maximum acceleration
amplitude the ramp was reversed to check for  an eventual
hysteresis.

In the following chapter 5 we describe in detail the type of
primary patterns, which appear near onset of the Faraday
instability. Secondary and higher order transitions towards more
complicated structures are dealt with in sections 6 and 7. Thereby
two representative experimental runs will be described in detail, the first
was taken at $f=6.25$ Hz$<f_b$ and a second at $f=7.25$ Hz$>f_b$.

In the former case the primary pattern exhibits a harmonic time
dependence, which turns out to be quite robust as it persists
over the whole investigated $\varepsilon$-ramp. The primary
spatial surface wave structure starts with  an ideal hexagonal
symmetry (region V), which then transforms into a pattern of squares
(region VI) as $\varepsilon$ is raised. This transition is
hysteretic, its global aspects can be understood in terms of a
simple model of six coupled amplitude equations.

In the second run at $f>f_b$ the primary surface pattern consists
of subharmonically oscillating squares (region II). On increasing
the drive strength $\varepsilon$ the interaction with the
neighboring harmonic Faraday instability leads to the appearance
of a quadratic $\sqrt{2} \times \sqrt{2}$ superlattice with a
displacive character in one and/or two lateral directions
(regions IIIa and IIIb, respectively). After crossing a phase
region of non-stationary patterns with a slow time dependence the
system enters a hexagonal $\sqrt{3} \times \sqrt{3}$ superlattice
(region IV). Mediated by a second local reconstruction process
the final stationary surface pattern is a quadratic $2 \times 2$
superlattice (region VII). Regardless of whether the response is s
or h, the surface finally breaks up and droplets are ejected
(region VIII, $\varepsilon \simeq 15-25\%$).

\section{Pattern formation close above onset of the Faraday instability}
Close to the harmonic onset of the instability ($f<f_b$) hexagons
are the preferred primary  surface pattern
(\ref{hexagonstosquares}a, region V)  for $f>f_b$, however,
squares are stable (figure~\ref{w2w2image}a). \cite {Wagner00}
(see  figure~3 in it) has shown that even for small $\varepsilon$
the wave profile is rather anharmonic.

\subsection{Theoretical model for the primary hexagonal pattern at $f<f_b$}
The appearance of hexagons at the ''harmonic side'' of the
bi-critical point follows from a triple wave vector resonance:
Since the spectrum of the harmonic Faraday mode $\zeta^h(t)$
consists of integral multiples of the drive frequency it allows
for two critical standing wave modes ${\bf k}_1$ and ${\bf k}_2$
to resonante with a third one. The requirement $|{\bf k}_1|=|{\bf
k}_2|=|{\bf k}_3|=k_c^h$ along with the resonance condition ${\bf
k}_1+{\bf k}_2+{\bf k}_3=0$ enforce a mutual angle of $120^\circ$
between the wave vectors implying the hexagonal symmetry. The
evolution equations for the respective mode amplitudes
$H_1,H_2,H_3$ are of the following structure
\begin{equation}
\partial_t H_1 = \varepsilon  H_1 + \beta
H_2^\star H_3^\star - \left [ |H_1|^2 +
\Gamma(120^\circ)(|H_2|^2+|H_3|^2) \right ]H_1.
\label{harmampl}
\end{equation}
Thereby  $\beta$ is a second order coupling coefficient and
$\Gamma(\theta)$ is the cubic cross coupling coefficient, which
depends on the angle between the  interacting modes. Moreover the
stars denote  complex conjugation. The corresponding equations
for $H_1$ and $H_2$ follow by permutation of the indices. The
term of cubic order is crucial for saturation. A linear stability
analysis of the finite amplitude stationary solution
$|H_1|=|H_2|=|H_3|$ given by \cite{{Ciliberto88}} yields a
backwards bifurcation out of the trivial solution $H_i=0$. This
reflects a hysteretic transition from the undisturbed flat
surface to a pattern of hexagons (compare
figure~\ref{bifdiagram}). However, we were unable to resolve any
hysteresis because of small amplitude (harmonically oscillating)
meniscus waves emitted from the rim of the container.

\subsection{Theoretical model for the primary square pattern at $f>f_b$}
Understanding the pattern selection process at the subharmonic
side of the bi-critical point is rather more complicated. Since the
frequency spectrum of the subharmonic Faraday response consist of
half integer multiples of $f$, any triple of linear unstable
modes is prevented from resonating. Thus nonlinear pattern
selection is dominated by a four-mode interaction. Unlike triad
resonances, which operate exclusively at an interaction angle of
$\theta=120^\circ$, four-wave resonances are less selective as
they work at arbitrary angles $\theta$. This fact is also
reflected by the corresponding system of amplitude equations.
Taking a set of $N$ standing waves with wave numbers ${\bf k}_i$
at length $|{\bf k}_i|=k_c^s$ but arbitrary relative orientation,
the respective mode amplitudes $S_i$ are governed by the
following evolution equations
\begin{equation}
\partial_t S_i = \varepsilon  S_i -
\sum^N_{j=1}\Gamma\left(\theta_{ij}\right)\left|S_j\right|^2S_i,
\label{subampl}
\end{equation}
with $\theta_{ij}$ being the angle between ${\bf k}_i$, ${\bf
k}_j$. Usually the participating modes are taken to be
equidistant on the circle $|{\bf k}_i|=k_c^s$, thus
$\theta_{i,i+1}=2 \pi/N$. In this case $N$ indicates the type of
symmetry of the pattern, namely $N=1$ lines, $N=2$ squares, $N=3$
triangles or hexagons, ... . As outlined in Refs.~(\cite
{Milner91}, \cite {Mueller94}, \cite {Zhang97} and \cite
{Chen97}) the question of what is the most preferred symmetry is
reduced to minimizing the ''free energy''
\begin{equation}
F=-\varepsilon \sum_{i=1}^N \left|S_i\right|+\frac{1}{2}
\sum_{i,j=1}^N \Gamma\left(\theta_{ij}\right) \left| S_i \right|^2
\left|S_j\right|^2
 \end{equation}
with respect to $N$ at given $\Gamma(\theta)$. For low viscosity
fluid layers of infinite depth the coupling function
$\Gamma(\theta)$ was first evaluated  by \cite {Zhang97}, who
found that a pattern of square symmetry is the most preferred one
at drive frequencies beyond $f \approx 50$Hz. At lower frequencies
patterns with a degree of rotational symmetry up $N=7$
(quasi-periodic) are likely to occur These predictions were
found to be in qualitative agreement with experiments
(\cite{Kudrolli96}, \cite{Binks97}). The latter authors also
extended the above considerations to the case of finite fill
heights and found square patterns to dominate also at lower drive
frequencies in agreement with their own experiments and also with ours.

\section{Secondary and higher order transitions at $f<f_b$}
\begin{figure}
\vspace{1pc} \caption{Photographs of the fluid surface as obtained
by ramping the drive amplitude $\varepsilon$ at $f=6.25$Hz. In the
corners of the pictures the container boundary is visible. In (a)
and (d) circles mark the nuclei at which the transition process
is initiated. (e) is an enlarged sector of (b) with marked
penta-lines and hepta-defects.} \label {hexagonstosquares}
\end{figure}
\subsection{Hexagon-square transition}
In this section we investigate the crossover from the primary
hexagonal structure (region V in figure~\ref{phasediagram}) to
the square pattern of region VI of the harmonic regime. Throughout the whole bifurcation
cascade the time dependence is purely harmonic without
perceptible subharmonic frequency contributions. The results
described below were obtained by ramping the drive amplitude
$\varepsilon =-0.1\%$ up to $\varepsilon=16.5\%$ while keeping
the frequency $f = 6.25 Hz$ fixed (see
figure~\ref{hexagonstosquares}). The transition is connected with
a strong hysteresis giving an overlap between the two ideal
structures.   Since the lateral aspect ratio (container size over
wavelength) is not large  the hexagonal pattern adapts to an
azimuthal symmetry. This affects the defect dynamic of the
pattern and thereby also the transition process. Starting the
$\varepsilon$-ramp at the hexagonal structure, a rearrangement of the pattern sets in, initiated by six nuclei of local quadratic order
as indicated by the circles in figure~\ref {hexagonstosquares}a.
The size of these patches increases with $\varepsilon$ (Fig.~\ref
{hexagonstosquares}b). Conversely, starting at elevated
$\varepsilon$ with the perfect square pattern  and ramping
downwards generates four nuclei of local hexagonal order
(Fig.~\ref {hexagonstosquares}d). Along the domain boundaries
between the patches penta-hepta defects occur. This kind of
defect is very  common in 2-D hexagonal patterns. Experimental
(\cite {Ciliberto88} and \cite {Tam00}) and theoretical (\cite
{Prismen93} and \cite {Tsimring96}) investigations reveal that a
penta hepta defect can be formed by a phase defect among two of
the participating modes. Figure~\ref{hexagonstosquares}e presents
an enlarged subrange of figure~\ref{hexagonstosquares}b depicting
"penta lines". This is a row of penta defects (unit cells with
five neighbors) ending in hepta defects (unit cells with seven
neighbors). Strengthening the drive makes the domain walls invade
the areas of hexagonal symmetry with new quadratic cells being
generated along the penta lines.

To quantify the hysteresis of the hexagonal quadratic
reconstruction we applied two different techniques, one in real
space and the other in Fourier space. In the former case the
number of unit cells with four neighbors ${\mathcal A}_4$ and
those with six neighbors ${\mathcal A}_6$ was counted.
 The result of this procedure is shown
\begin{figure}
   \vspace{1pc}
\caption{(a) Surface area covered with squares ${\mathcal A}_4$
related to surface area of squares and hexagons ${\mathcal A}_6$
as  function of the drive strength $\varepsilon$ at the frequency
$f=6.25$ Hz. (b)  The autocorrelation function $C\left(\phi =
90^\circ \right)$ of the power spectra $P({\bf k})$ on the circle
($|k| = k_c$). Upright (reversed) triangles refer to the upwards
(downwards) amplitude ramp.}
  \label {area}
\end{figure}
in figure~\ref{area}a. It  reveals a hysteresis loop extending
from  $\varepsilon = 5\%\pm 1\%$ up to $\varepsilon =15\%\pm1\%$.
The obvious staircase behavior reflects  the discretization of the k-values due to the finite size of the container.
For different runs the steps do not occur at the same
$\varepsilon$-values.

The second method for evaluating the square-vs.-hexagon surface
coverage we Fourier transformed the full surface picture and
evaluated the power spectrum $P({\bf k})=P(k,\phi)$ at $|{\bf
k}|=k_c$ as a function of the azimuthal angle $\phi$. We then
computed the azimuthal auto correlation $C(\phi)$ as follows
\begin{equation}
C(\phi)=\frac{\int_0^\pi P(k_c,\phi)P(k_c,\phi+\phi') {\rm d}
\phi'}
 {\int_0^\pi [P(k_c,\phi')]^2 {\rm d} \phi'}.
  \end{equation}
That way a pattern with quadratic symmetry leads to a peak at
$\phi=90^\circ$ and $180^\circ$, while a hexagonal pattern
produces maxima at $\phi=60^\circ$, $120^\circ$ and $180^\circ$.
\begin{figure}
   \vspace{1pc}
  \caption{Azimuthal autocorrelation $C(\phi)$ derived from the power
   spectra $P({\bf k})$ at the wave number $|{\bf k}| = k_c$ as a
   function of the drive amplitude $\varepsilon$ for ramping up and down.}
  \label {autocorr}
\end{figure}
Figure~\ref{autocorr} illustrates how $C(\phi)$ develops with the
drive amplitude, both at increasing and decreasing $\varepsilon$.
In order to compare with the  cell counting method (see
figure~\ref {area}a) the value $C(\phi=90^\circ)$ is plotted
versus $\varepsilon$ in figure~\ref {area}b. Regarding the width
of the hysteresis loop, the two methods agree within a few percent
with each other.

An important feature of the above phase transition is the
constancy of the wave number $k^h(\varepsilon)=k^h_c$: Within the
experimental resolution of $\Delta k/k=\pm 1\%$ no dependency of
the wave number on $\varepsilon$ could be detected throughout the
whole investigated drive amplitude range. This is of particular
significance as it allows for to describe the global aspects of
the phase transition  in terms of a simple model as given in the
next section.

\subsection{Comparison with theory}
It was outlined in ~\S\,5 that the hexagonal symmetry at small
$\varepsilon$ is a consequence of a three wave resonance,
reflected by the second order term  in equation (\ref{harmampl}).
However, upon
increasing the control parameter $\varepsilon$ the term of cubic
order becomes increasingly important thus finally enforcing the
transition towards squares.  A minimal model that allows
stationary solutions in form of squares and hexagons (and also
lines) along with a linear stability analysis of these solutions
can be found in  a recent publication of \cite {Regnier97}. This
model is an extension  of equation (\ref{harmampl}) as it relies
on six independent modes ${\bf k}_{i, i=1...6}$ rather than only
three (see figure~\ref {sixwavevectors}).
\begin{figure}
   \vspace{1pc}
  \caption{Set of wave vectors necessary to build up patterns of
  rolls, hexagons or squares.}
  \label {sixwavevectors}
\end{figure}
Since ${\bf k}_{i, i=1,2,3}$ enclose mutual angles of $120^\circ$ the
remaining  wave vectors ${\bf k}_{i, i=4,5,6}$  describe a second
pattern of hexagons, which is rotated by an angle of $30^\circ$
relative to the first. Depending on the number of saturated mode
amplitudes $H_i$ the model allows stationary solutions in form of
lines, squares, hexagons or even more complicated solutions
without translational symmetries (quasi-periodic structures). We
shall disregard these complications here and focus on the
square-hexagon competition. \cite{Regnier97} demonstrate that the
hexagonal state (given by finite $|H_1|=|H_2|=|H_3|$) becomes
unstable against shear distortions $\delta$ in form of rhombuses ($\delta H_2
\ne 0 \ne \delta H_3$) at a drive amplitude
$\varepsilon>\epsilon_{H}$.
By way of contrast, for a critical value
of $\varepsilon<\epsilon_{S}$ squares (given by finite
$|H_1|=|H_4|$) become unstable against rhomboedric ($H_2$, $H_3$)
disturbances. These results are summarized in the bifurcation
diagram of figure~\ref {bifdiagram}.

%
\begin{figure}
\vspace{1pc} \caption{Sketch of the bifurcation diagram for the
transition from a hexagonal to a quadratic pattern. Dotted lines
indicate unstable, straight lines stable branches. The hysteresis
observed in the experiment may be attributed to the bi-stable
region, where squares and hexagons co-exist. } \label {bifdiagram}
\end{figure}
It is tempting to attribute the bi-stable square-hexagon region
between $\varepsilon_S < \varepsilon < \varepsilon_H$ (see
figure~\ref {bifdiagram}) to the hysteretic region depicted in
figure~\ref {hexagonstosquares}b,c). Note however, that the
observed transition runs through a reconstruction via penta-hepta
defects, the complicated space dependence of which goes beyond the
scope of the present model.

We mention that a discussion of a transition between hexagons and
lines in terms of the above three mode model (\ref {harmampl}) has
been given earlier in the context of Rayleigh-Benard convection
by \cite {Walden81} and \cite {Ciliberto88}. Thereby the phase
transition results from non-Boussinesq effects induced by a
strong applied temperature gradient. Recently, several authors
report B\'{e}nard-Marangoni experiments, which show a transition from
hexagons to squares (\cite {Nitschke95}, \cite {Bestehorn96} and
\cite {Eckert98}). A similar transition is found by \cite
{Abou00} on the surface instrability of magnetic liquids
(Rosensweig-Instability). However, unlike our measurements, all
of the afore mentioned experiments with a hexagon to square
transition exhibit a considerable
nonlinear wave number variation $k(\varepsilon)$ up to  $10\%$,
which rules out a description in terms of space independent
amplitude equations of the type given above.

\section{Secondary and higher transitions at $f>f_b$}
In the same way as in the preceding paragraph we now turn to the
bifurcation scenario at the opposite side of the bi-critical
point at $f>f_b$. An amplitude ramp taken at $f=7.25Hz$ serves as
a representative example.  ~\S\,7.4 is an exception, where the
focus is on the frequency regime $6.6 Hz < f < 6.9 Hz$.

\begin{figure}
\vspace{1pc} \caption{Phase diagram of an amplitude scan at $f =
7.25$ Hz. Time dependent transient means an uncorrelated pattern
(figure~\ref{transient}), quasistationary indicates a pattern of
an almost perfect $\sqrt 3 \times \sqrt 3$ superlattice (compare
figure~\ref {w3w3image}) with a slow defect dynamic.}
\label{updownramp}
\end{figure}
An overview over the transition behaviour at $f=7.25Hz$ is given
by figures~\ref{updownramp} and \ref{w2w2image}. Starting from
subharmonic oscillating squares (region II,
figure~\ref{w2w2image}a)
a transition to a quadratic $\sqrt2
\times \sqrt 2$ superlattice with a displacement of neighbouring
elevation maxima in the lateral $x$-direction (region IIIa,
figure~\ref{w2w2image}c) takes place. The next pattern is again a $\sqrt2
\times \sqrt 2$ superlattice of the original square lattice but this time it
exhibits a displacement in both $x$
and $y$ direction.
(region IIIb, figure \ref{w2w2image}e). After a
time dependent transient (see figure~\ref{transient} for a
snapshot) this pattern transforms into a "quasi-stationary"
hexagonal $\sqrt 3 \times \sqrt 3$ superlattice (region IV,
figure~\ref{w3w3image}). With "quasi-stationary" we indicate that
the pattern is slightly disturbed by defects, which induce a slow
time dependence on the scale of minutes.
Further raising the drive amplitude
$\varepsilon$ at $f=7.25$Hz makes the
quadratic symmetry reappear in form of a $2 \times 2$ superlattice
(region VII, figure~\ref{2x2image}). Performing the same
amplitude ramp at a lower frequency of $6.6 Hz < f < 6.9$Hz the
$\sqrt 3 \times \sqrt 3$ superlattice directly reduces to its
underlying pure hexagonal tiling. This last transition will be
discussed in ~\S\,7.4.

\subsection{The quadratic $\sqrt2 \times \sqrt 2$ superlattice}
\begin{figure}
\vspace{1pc} \caption{A $10 \times 10$ cm sector of the
photographs of the fluid surface and corresponding Fourier spectra
at $f = 7.25$ Hz. a,b): $\varepsilon = 3\%$, region II, pattern
of squares;  c,d): $\varepsilon = 6.6\%$, region IIIa, $\sqrt 2
\times \sqrt 2$ superlattice p2mg with a displacive character in
$x$-direction; e,f): $\varepsilon = 14\%$ $\sqrt 2 \times \sqrt
2$, region IIIb, superlattice c2mm with a  dispalzive character
in $x$- and $y$-direction. The dotted lines and arrows on the
left pictures indicate the directions in which the rows and
columns of elevation maxima are displaced.} \label {w2w2image}
\end{figure}
\begin{figure}
\vspace{1pc} \caption{Particle model of the a) $\sqrt 2 \times
\sqrt 2$ p2mg, b)$\sqrt 2 \times \sqrt 2$ c2mm superlattice. Also
marked are the mirror (m) and glide (g) planes.} \label
{particelmodel}
\end{figure}
The bifurcation sequence starts at the primary ideal pattern of
subharmonically oscillating squares (region II and
figure~\ref{w2w2image}a,b.) composed of the two fundamental wave
vectors ${\bf k}_{S1}$ and ${\bf k}_{S2}$. Increasing the drive
strength $\varepsilon$ beyond $5\%$ displaces every other column
of elevation maxima in the direction $\pm({\bf k}_{S1}+{\bf
k}_{S2})$. This is shown by the arrows in
figure~\ref{w2w2image}a. The resulting pattern (phase region
IIIa) is depicted in \ref{w2w2image}c.  The displacement is
accompanied by the simultaneous appearance of the modes ${\bf
k}_{H1}$, ${\bf k}_{H2}$, and  ${\bf k}_{D1}$ in the power
spectrum (see figure~\ref{w2w2image}d). Due to the approximate
equality $|{\bf k}_{H1}|=|{\bf k}_{H_2}|\simeq k_c^h$ these modes
are only slightly damped. In contrast, $|{\bf k}_{D1}|$ is
significantly smaller than  $k_c^s$and $k_c^h$. Therefore the
mode $D_1$ is strongly damped. Its mission is to act as a
mediator mode, enabling the resonance between $S$- and $H$-modes
according to the geometrical rules ${\bf k}_{H_1}= \left({\bf
k}_{S_1}+{\bf k}_{D_1}\right)$ and ${\bf k}_{H_2}= \left({\bf
k}_{S_2}+{\bf k}_{D_1}\right)$. This implies that ${\bf k}_{D1}$
must be associated with a subharmonic time dependence. Moreover,
it follows from $k_c^h/k_c^s\simeq 1.58$ that $|{\bf
k}_{D1}|=0.71\simeq 1/\sqrt{2}$, which leads to the commensurate
relationship ${\bf k}_{D1}\simeq \frac{1}{2}({\bf k}_{S_1}+{\bf
k}_{S_2})$ for the new fundamental wave vector. The corresponding
elementary cell (see figure~\ref {particelmodel}a) is rotated by
$45^\circ$ relative to the original quadratic grid, and the basic
wavelength is enlarged by a factor of $\sqrt 2$. We therefore
denote  this pattern as a $\sqrt 2 \times \sqrt 2$ superlattice
(more precisely: ($\sqrt 2 \times \sqrt 2)R45$ or $\sqrt 2 \times
\sqrt 2$ $p2mg$ in the nomenclature of space group theory). It
turns out that the displazive character of the superlattice is
inherent in the phase information carried by the participating
modes. Barring higher spatial harmonics the space dependence of
the surface deformation shown in figure~\ref{w2w2image} can be
expressed as
\begin{equation}
\eta({\bf r}) = \sum_{i=1}^2 S_i \cos\left({\bf k}_{S_i}{\bf
r}+\phi_{S_i}\right) + D_1 \cos\left({\bf k}_{D_1}{\bf
r}+\phi_{D_1}\right) + \sum_{i=1}^2 H_i \cos\left({\bf
k}_{H_i}{\bf r}+\phi_{H_i}\right). \label {surface}
  \end{equation}
By means of the ray tracing technique outlined in ~\S\,2.2 we can
now simulate the video image associated with $\eta({\bf r})$ and
adapt it to the empiric result. Taking $\phi_{S_1}=\phi_{S_2}=0$
(by a proper choice of the origin) our investigation reveals that
the displacement visible in figure~\ref {w2w2image}b can only be
reproduced if the spatial phases $\phi_{D_1}$, $\phi_{H_1}$, and
$\phi_{H_2}$ adopt values close to $=\pi/2$. The associated
surface pattern exhibits a $180^\circ$ rotational symmetry.

By increasing the drive further the $\sqrt 2 \times \sqrt 2$
$p2mg$ superlattice  undergoes a transition which restores the
fourfold symmetry. Similar to the above described shift of the
{\em columns} of elevation maxima, it is now additionally the
{\em rows}, which experience a displacement in the direction $\pm
({\bf k}_{S_1}-{\bf k}_{S_2})$ (indicated by the arrows in
figure~\ref{w2w2image}c). In figure~\ref{D1D2} we measured the
amount of symmetry restoration by comparing the spectral power
associated with $D_1$ and $D_2$. Beyond $\varepsilon \simeq 9\%$
the transition is complete. The resulting so called $\sqrt 2
\times \sqrt 2$ $c2mm$ superlattice (see figure~\ref
{particelmodel}b) is depicted in figure~\ref {w2w2image}e and
associated with the phase space region IIIb. In Fourier space
this transition is carried by the additional modes ${\bf
k}_{H_3}$, ${\bf k}_{H_4}$ and ${\bf k}_{D_2}$ as shown in
figure~\ref {w2w2image}f. Extending the surface
representation~\ref{surface} by these additional components and
using it to re-perform the ray tracing image analysis yields the
phase information $\phi_{D_{i,
i=1,2}}=\phi_{H_{i,i=1,..,4}}=\pi/2$.

We mention that the  $\sqrt 2 \times \sqrt 2$ $c2mm$ superlattice
state was not observed in an earlier measurement on a liquid of
higher viscosity (\cite{Wagner00}). Thereby this structure is preempted
by a transition to a hexagonal symmetry.

\subsection{Theoretical model for the displacive phase}
The experimental investigations outlined in the previous section
reveal that the prominent displacive character of the $\sqrt2
\times \sqrt 2$ superlattices is associated with the phase
information carried by the spatial Fourier modes. In what follows,
a minimal model is constructed, which is able to explain the
experimentally measured phases. It is important to point out that
the {\em structure} of these equations just relies on symmetry
and resonance arguments (triad wave vector resonances). The  {\em
numerical values} of the appearing coefficients are not known;
their evaluation would require a rather complicated nonlinear
analysis. Furthermore, for the sake of simplicity, we limit our
discussion to the $\sqrt2 \times \sqrt 2$ $p2mg$ superlattice.
The generalization to the more symmetric $\sqrt2 \times \sqrt 2$
$c2mm$ pattern is straightforward.

Assuming that the amplitudes of the primary square pattern have settled at some
finite value $S_1=S_2\ne0$ the leading order behavior of the
remaining modes is governed by the following set of equations
\begin{eqnarray}
\partial_t D_1 & = & \varepsilon_D  D_1 + \mu_D
\left(S_1^\star H_1+ S_2^\star H_2 \right)+ \chi_S S_1 S_2
D_1^\star \nonumber\\
\partial_t H_1 & = & \varepsilon_H H_1 + \mu_H S_1 D_1
\\
\partial_t H_2 & = & \varepsilon_H H_2 + \mu_H S_2 D_1
. \nonumber
 \label {w2w2}
\end{eqnarray}
Thereby $\varepsilon_H<0$ and $\varepsilon_D\ll 0$ are the
coefficients of linear damping of respectively the $H$ and $D$
modes, while  $\mu_{D,H}$ and $\chi_{S}$ are nonlinear coupling
coefficients associated with the triad wave vector resonance.
\begin{figure}
   \vspace{1pc}
  \caption{Integrated intensity of the $D_1$-peak  (circles) and the
  $D_2$-peak
   (squares) in the power spectrum as a function of the drive
   amplitude $\varepsilon$. The error bars mark
the standard derivation resulting from 5 succeeding runs.}
  \label {D1D2}
\end{figure}

Although  the model equations (\ref{w2w2}) are linear in $D$ and $H$ and thus
saturation is not implied, the appearing nonlinearities are phase
selective. By writing the complex amplitudes in the form
$A_i=|A_i|\exp\left({\phi_{A_i}}\right)$ with again
$\phi_{S1}=\phi_{S2}$ taken to be zero (choice of space origin)
the imaginary part of equation (\ref{w2w2}) yields the phase
dynamics
\begin{eqnarray}
\partial_t \phi_{D_1} & = & \mu_D |S||H|/|D|
\left[\sin\left(\phi_{H_1}-\phi_{D_1}\right)+
\sin\left(\phi_{H_2}-\phi_{D_1}\right)\right]+ \chi_S
\left|S\right|^2\sin\left(-2\phi_{D_1}\right) \nonumber \\
\partial_t \phi_{H_1} & = & \mu_H |D||S|/|H|
\sin\left(\phi_{D_1}-\phi_{H_1}\right)   \\
\partial_t \phi_{H_2} & = & \mu_H |D||S|/|H|
\sin\left(\phi_{D_1}-\phi_{H_2}\right).\nonumber
 \label {w2w2phase}
\end{eqnarray}
 The fix points of these equations are
\begin{equation} \phi_{H_i}=\phi_{D_i}=m \frac{\pi}{2}
\end{equation}
with $m$ being an integer. By inspection one finds that the
solution with even $m$ leads to an square pattern with {\em
amplitude modulation} while odd $m$ gives rise to the observed
displacement ({\em phase modulation}). Which one is stable,
depends on the numeric values of the coefficients. From the
experimental data we conclude that $\phi_{H_i}=\phi_{D_i}= \pi/2$
is the solution applicable to the present experiment.

\subsection{The hexagonal $\sqrt3 \times \sqrt 3$ superlattice}
\begin{figure}
   \vspace{1pc}
  \caption{Snapshot of the fluid surface as obtained at
$f=7.25$ Hz and  $\varepsilon = 17\%$. The pattern is strongly
time dependent and spatially uncorrelated.}
  \label {transient}
\end{figure}
\begin{figure}
\vspace{1pc} \caption{A $8 \times 8$ cm sector of the photographs
of the fluid surface and corresponding Fourier spectra at $f =
6.65$ Hz and $\varepsilon = 5 \%$. This photograph has been
obtained by using a diffusive light source mounted off axis
instead of the standard visualization technique. That way the
character of the hexagonal $\sqrt 3 \times \sqrt 3$ superlattice
(region IV) appears to be more pronounced (compare figure~4e in
(Wagner (2000))).} \label {w3w3image}
\end{figure}
Upon further increase of the driving force at $f=7.25Hz$ the
quadratic $\sqrt2 \times \sqrt 2$ $c2mm$ superlattice transforms
into a "quasi-stationary" hexagonal superlattice (region IV,
compare figure~\ref{w3w3image}) after passing a region of
transient time dependent (figure~\ref {transient}) with squares,
hexagons and disordered patterns appearing at the same value of
$\varepsilon$.  The term "quasi-stationary" is to indicate that
the pattern is affected by defects on a slow time scale of
minutes. The power spectrum shown in Figure~\ref{w3w3image}b
reveals that the structure is composed of a set of three harmonic
modes $H_j$ and three subharmonic modes $S_j$. The angles between
the wave vectors of the three  $H$ modes is $120^\circ$ and so is
the angle of the S-modes. The H-pattern is rotated by $30^\circ$
and locked in phase with respect to the S-pattern, the ratio
$k_h/k_s$ is $1.73 \simeq \sqrt3$. Therefore the pattern is
denoted as a $\sqrt3 \times\sqrt 3$ superlattice. We mention an
earlier investigation on a more viscous fluid of $\nu=10$cS
(\cite{Wagner00}): Thereby the transition from the quadratic
$\sqrt2 \times \sqrt 2$ $p2mg$ to the hexagonal $\sqrt3 \times
\sqrt 3$ superlattice took place in a more correlated manner via
stacking faults while the $\sqrt2 \times \sqrt 2$ $c2mm$ state
could not be observed.

At higher $\varepsilon$-values the further development
of the $\sqrt3 \times \sqrt 3$ superlattice depends on the drive
frequency: At $6.6$Hz$<f<6.9$Hz, where the superlattice is
perfectly stationary and almost free of defects, the amplitude of
the subharmonic components of the structure continuously become smaller
until the hexagonal base pattern with a pure synchronous time dependence
remains (region V, compare figure~\ref {hexagonstosquares}a).
This trannsition  is slightly hysteretic ($\Delta \varepsilon \simeq
1-2\%$) and can be localized accurately by means of the  triggering
technique described in ~\S\,2. We come back to this crossover in
\S\,7.4, in order to present a theoretical model.

At drive frequencies larger than $\approx 6.9$Hz (for concreteness
let us return to our run at $f=7.25$Hz) the hexagonal $\sqrt3
\times \sqrt 3$ superstructure transforms into a  $2 \times 2$
superlattice (region IV $\rightarrow$ VII)
thereby restoring the quadratic symmetry. The crossover takes place via a
spatially weakly correlated transient, being subjected to a
strong temporal dynamic. The corresponding pattern looks similar to
the surface state shown in figure~\ref{transient}. In its
final state the subharmonic and the harmonic wave vectors ${\bf
k}_{S,H}$ are aligned with each other (figure~\ref {2x2image}b).
With $k^h \approx k_c^h$ but $k^s\approx 0.8k_c^s$ the resulting
length ratio $k_H/k_S$  is about 2.

\subsection{Theoretical model for the $\sqrt3 \times \sqrt 3$
superlattice-to-hexagon transition}

The following section provides a theoretical model for the
transition from the $\sqrt3 \times \sqrt 3$ superlattice (region
IV) to the pure hexagon state in region V, taking place at $6.6$Hz
$<f<6.9$Hz. Following the lines of  ~\S\,7.2 spatial and temporal
resonance arguments are used to construct the set of amplitude
equations. We consider the underlying hexagonal base pattern,
carried by the modes  $H_i$ (c.f. equation (\ref{harmampl})) as
being saturated at some finite amplitude. Then the bifurcation of
the subharmonic modes $S_1$, $S_2$, $S_3$ is governed by the
following system of equations
\begin{equation}
\partial_t S_1 = \varepsilon  S_1 + \beta_{HS}
\left(H_1 S_3+ H_2^\star S_2 \right)-\mu
\sum^3_{j=1} \left|H_j\right|^2 S_1 +h.o.t. . \label {w3w3}
\end{equation}
where the equations for $S_2$ and $S_3$ follow by cyclic
permutation. and $\beta_{HS}$ and $\mu$ are nonlinear coupling
coefficients. In this model the  resonance condition between the
$S$ and the $H$ modes enforces a wave number ratio
$k_H/k_S|_{theo} = \sqrt3 \approx 1.73$. Our measurements reveal
$|{\bf k}_h|= k^H_c(\varepsilon=4\%)$ and $|{\bf k}_s|=0.91
k_c^S$ giving $k_H/k_S|_{exp} \approx 1.73$. This justifies our
assumption that harmonic modes dominate the pattern forming
process, while the (slaved) subharmonic ones have to adapt their
wavelength.

The occurrence of the $\sqrt3 \times \sqrt 3$ superlattice is
caught by a linear stability analysis of equation (\ref{w3w3}). To
that end it is convenient to take the amplitude
$H=|H_1|=|H_2|=|H_3|$ as the control parameter.  Then one obtains
a positive growth rate (instability) for the $S_i$ if
\begin{equation}
\varepsilon + 2 \beta_{HS}|H| - \mu |H|^2 > 0.
 \label {w3w3stability}
\end{equation}
This condition determines the transition to the superlattice.
Recall that (\ref{w3w3}) does not entail terms which are
nonlinear in $S_i$ thus the model neither explains saturation of
the $S_i$ nor the experimentally observed hysteresis in the
transition.

Let us now focus on  the spatial phase dynamic of the hexagonal
superlattice. For the underlying hexagonal structure with
amplitudes $H_i=|H_i| \exp{(i \Phi_{H_i} )}$ we have either
$\Phi_{H} = \sum_j^3 \phi_{H_j}=0$ or $\pi$, corresponding
respectively to ''up'' or ''down'' hexagons. For Faraday waves
this distinction is not significant because the surface
oscillation periodically switches between these two possible
states. Splitting the subharmonic mode amplitudes also into
modulus and phase and assuming without loss of generality that
$\phi_{H_1}=\phi_{H_2}=0$, the  imaginary part of equation
(\ref{w3w3}) yields the identity
$\phi_{S_1}=\phi_{S_2}=\phi_{S_3}$. In order to determine that
value it is necessary to proceed with the amplitude expansion up
to the quintic order. Barring all terms which are not phase
selective the extension of equation (\ref{w3w3}) reads
\begin{equation}
\partial_t S_1 = \varepsilon  S_1 + \beta_{HS}
\left(H_1 S_3+ H_2^\star S_2 \right)+ .. + \chi S_1^\star
\left(S_2^\star\right)^2\left(S_3^\star\right)^2 + ...,
 \label {w3w3quintic}
\end{equation}
with $\chi$ being a nonlinear coefficient. Introducing the
parametrization $\Phi_S=\sum_j^3 \phi_{S_j}=3\phi_{S_i}$ equation
(\ref{w3w3quintic}) yields the stationary solution
\begin{equation}
\Phi_S=n \pi/2.
\end{equation}
where $n$ is an an integer. Following the usual
nomenclature the pattern with odd $n$ is a triangular
superlattice, as its rotational symmetry is threefold. For even $n$
the superlattice exhibits a sixfold symmetry
(c.f.~\cite{Mueller93}). Since a sixfold symmetry center can be
easily identified in figure~\ref{w3w3image}, we conclude that $n$
is even in our experiment. We point out that an example of a
$\sqrt 3 \times \sqrt 3$ with third order rotational symmetry
($n$ is odd) has recently been observed by \cite{Pi00} (see also
\cite{Wagner01}).

\begin{figure}
\vspace{1pc} \caption{A $8 \times 8$ cm sector of the quadratic $2
\times 2$
 superlattice and corresponding Fourier spectra at $f= 7.25
 Hz$ and $\varepsilon  = 24\%$ (regionVII).}
\label {2x2image}
\end{figure}

\section{Conclusions}

We have presented a comprehensive investigation on Faraday wave
pattern formation in the vicinity of a bi-critical situation.
Thereby modes with harmonic and subharmonic time dependence
interact and lead to a variety of superlattice states. This kind
of structure acts as a mediator during the transition process
between two incompatible space groups (squares and hexagons). We
developed several sets of amplitude equations to model the
observed phase transitions. Special attention is devoted to the
phase information carried by the participating modes, which is
responsible for the  remarkable displacive feature of one of the
superlattices.

Superlattices are rather common in 2-D solid state physics, and a
comparison is therefore instructive. The transition from a simple
hexagonal lattice to a $\sqrt{3} \times  \sqrt{3}$ superstructure
has been observed for instance in monolayers of $C_2ClF_5$
adsorbed on graphite  (\cite{Fassbender95}). The transition from
the subharmonic quadratic base pattern to a $\sqrt{2} \times
\sqrt{2}$ superlattices with a displacive character in one
direction ($p2mg$, region IIIa) is analogous to the
reconstruction of the clean (100) surface of W (=tungsten) crystals
(\cite{Schmidt92}). Here the surface atoms are displaced in
exactly the same way as the elevation maxima of the surface
profile in the present study. Most interestingly, the surface of
W crystals obeys a transition to a $\sqrt{2} \times
 \sqrt{2}$
superlattices with a displacive character in two directions
($c2mm$, identical to our pattern in region IIIb)
 in the presence of Hydrogen atoms.

We were also able to provide for the first time  a pattern forming
system, which undergoes a hexagon-to-square transition without a
simultaneous change of the fundamental wavelength.
This is in contrast to earlier observations on
the B\'{e}nard Marangoni system, which showed considerable wave
number variations during the transition process. The accompanying
description in terms of well defined spatial Fourier modes, makes
the phase transition very fundamental and identifies the
underlying pattern selection mechanism to switch from a triad to
a four-wave vector resonance.

\begin{acknowledgments}
We thank J.~Albers for his support. This work is supported by the
Deutsche Forschungsgemeinschaft.
\end{acknowledgments}

\end{document}